%% file: main.tex
\documentclass[conference]{IEEEtran}
\IEEEoverridecommandlockouts

\usepackage{cite}
\usepackage{amsmath,amssymb,amsfonts}
\usepackage{algorithmic}
\usepackage{graphicx}
\usepackage{textcomp}
\usepackage{xcolor}
\usepackage{subcaption}
\usepackage{booktabs}
\usepackage{caption}
\usepackage[english]{babel}
\usepackage{csquotes}
\usepackage{comment}
\usepackage{tabularx}
\usepackage{todonotes}
\usepackage{pgfplots}
\usepackage{pgfplotstable}
\usepackage{placeins}
\graphicspath{{./svg-inkscape/}}
\usepackage{hyperref}
\hypersetup{colorlinks=true, urlcolor=blue}
\usepackage{cleveref}
\definecolor{kahypar}{HTML}{2196F3}   
\definecolor{greedy}{HTML}{4CAF50}    
\definecolor{zoltan}{HTML}{FF9800}    
\definecolor{fm}{HTML}{9C27B0}        
\definecolor{ea}{HTML}{E91E63}        
\definecolor{stocg}{HTML}{00BCD4}     
\definecolor{randomp}{HTML}{9E9E9E}   

\def\BibTeX{{\rm B\kern-.05em{\sc i\kern-.025em b}\kern-.08em
    T\kern-.1667em\lower.7ex\hbox{E}\kern-.125emX}}
\begin{document}

\title{On the Distortion of Partitioning Performance by Random Quantum Circuits \\ 
\thanks{This research was supported by the EPSRC UK Quantum Technologies Programme under grant EP/T001062/1
and VeriQloud.}
}

\author{\IEEEauthorblockN{1\textsuperscript{st} Maria Gragera Garces}
\IEEEauthorblockA{\textit{Quantum Software Lab, University of Edinburgh} \\
Scotland, UK \\
0009-0000-9018-7435}
}

\maketitle

\vspace{-0.8 cm}

\begin{center}
{\footnotesize\textit{Accepted at ICDCS2026 (DisQIC)}}
\end{center}

\begin{abstract}
Hypergraph partitioning is a central component of distributed quantum computing (DQC) compilers.
However, due to the limited size of available quantum benchmark suites, many partitioning studies rely on random quantum circuits as evaluation workloads.
In this work, we investigate whether such benchmarking practices provide a faithful assessment of partitioner performance.

We evaluate a diverse set of state-of-the-art hypergraph partitioning strategies across three circuit origins: real algorithmic circuits, structured generated circuits, and fully random circuits.

Our results show that random circuits significantly distort partitioning evaluation.
They inflate cut costs, alter scaling trends across QPU counts and circuit sizes, and change the relative ranking of partitioning strategies.
In contrast, structured generated circuits exhibit substantially lower distortion, more closely approximating real workload behaviour in cost, scaling, and strategy rankings.
These findings demonstrate that benchmark selection directly influences methodological conclusions in DQC research and that random circuits may provide misleading guidance for compiler design.
\end{abstract}
\vspace{-0.3cm}

\section{Introduction}
Distributed Quantum Computing (DQC) is widely regarded as one of the most promising paths toward scalability in the near and long term \cite{cuomo2020towards}.
Instead of relying on a single monolithic quantum processor, DQC distributes a large computational workload across a network of smaller quantum processing units (QPUs) that collaboratively execute the task.
This modular paradigm aims to enable large-scale computations beyond the physical limitations of individual devices.

A central compiler-level challenge in DQC is how to divide a quantum computation into sub-computations that can be executed across multiple QPUs while minimizing communication overhead.
The standard abstraction for this problem is the hypergraph representation of a quantum circuit \cite{barral2025review}, where qubits correspond to vertices and multi-qubit gates induce hyperedges.
Partitioning this hypergraph yields subcircuits that are mapped to different devices, while cut hyperedges correspond to required quantum communication primitives such as teleportation or entanglement distribution \cite{andres2019automated}.

Hypergraph partitioning algorithms therefore form the backbone of many DQC compilation pipelines.
A wide spectrum of methods is available, ranging from greedy heuristics to sophisticated multilevel partitioners such as KaHyPar \cite{schlag2023high}.
Despite their algorithmic differences, these methods share a common principle: they exploit the structural properties and interconnectivity patterns of the hypergraph.
The quality of the partitioning is therefore inherently tied to the shape and topology of the underlying circuit-derived hypergraph.

However, much of the current benchmarking literature evaluates partitioning strategies using random quantum circuits as test instances 
\cite{sundaram2023distributing,sundaram2021efficient,burt2026multilevel,escofet2023hungarian,bandic2023mapping}. 
While random circuits are convenient and used for performance evaluation \cite{arute2019quantum}, 
they possess structural characteristics that may differ significantly from those of practical or algorithmic circuits \cite{fisher2023random}.

In this work, we investigate how the origin of circuit-derived hypergraphs 
(whether from "real" circuits resulting from well-known quantum algorithms, 
circuits generated by generative models, 
or fully random circuits)
influences partitioning performance.
We compare partitioners not only against each other under standard benchmarks, but also against a fully random assignment baseline.
To formalize this concern, we define \emph{distortion} as a measurable deviation in partitioner evaluation outcomes induced by the choice of circuit ensemble.
Distortion is not binary but a matter of degree: an ensemble is more distorted if it departs further from real workloads in cut cost, scaling behaviour,
strategy ranking, 
or result variability. 

Our results demonstrate that random circuits can distort the perceived quality of a hypergraph partitioner.
They may overestimate or underestimate algorithmic advantage by masking the structural features that real workloads exhibit.
This has important implications for how distributed quantum compilation strategies are evaluated.

\vspace{-0.25 cm}
\section{Methodology}
\vspace{-0.18 cm}

To distribute quantum circuits, we abstract them to hypergraphs, where:
\begin{itemize}
\item vertices correspond to qubits,
\item hyperedges correspond to multi-qubit gates.
\end{itemize}

Partitioning this hypergraph assigns sub-circuits to QPUs; cut hyperedges correspond to
non-local operations realised via quantum communication primitives \cite{andres2019automated}.
We evaluate a broad spectrum of partitioning strategies in this setting \cite{ccatalyurek2023more,barral2025review},
described below.

\subsection{Partitioners}
We evaluate the following hypergraph partitioning strategies, ordered by algorithmic sophistication.
All hypergraphs are balanced under the constraint
\[
\text{max\_block} = \left\lfloor \frac{n}{k}(1+\epsilon) \right\rfloor + 1,
\]
where $n$ is the number of vertices (qubits), 
$k$ the number of partitions (QPUs accross which the circuits will be distributed), 
and $\epsilon$ the imbalance tolerance (set to $5\%$).

\subsubsection{Random partitioner}
Our random partitioner assigns each vertex independently and uniformly
to one of the $k$ blocks.
No structural information from the hypergraph is used.
The resulting cut size serves as a baseline
for evaluating the benefit of all other methods.

\subsubsection{Greedy partitioner}
Our greedy partitioner processes hyperedges sequentially.
For each hyperedge, unassigned vertices are placed in the block
that already contains the majority of that hyperedge's vertices.
Ties are broken by selecting the least-loaded block.
If the preferred block exceeds capacity,
vertices are assigned to the least-loaded feasible block.
This method is deterministic, single-pass, and computationally inexpensive,
but performs no refinement or backtracking.

\subsubsection{Stochastic Greedy partitioner (StochG)}
StochG extends the greedy approach with randomised restarts
and local improvement.
Each iteration randomly shuffles hyperedge order,
performs greedy construction,
and then applies a local improvement sweep
where vertices are moved only if the move strictly reduces cut cost
while respecting balance constraints.
The algorithm runs within a fixed time budget
and returns the best solution found.
Randomisation allows exploration of multiple local minima.

\subsubsection{Fiduccia-Mattheyses (FM)}
We implement a $k$-way extension of the originical
Fiduccia--Mattheyses hypergraph partitioner \cite{fiduccia1988linear}.
The algorithm begins from a balanced initial assignment.
During each pass, it computes the gain of moving each unlocked vertex
to alternative blocks and greedily applies the highest-gain legal move.
Vertices are locked once moved within a pass.
At the end of each pass, the assignment is rolled back
to the best state encountered.
Passes repeat until no further improvement is observed.
FM provides structured local refinement
and typically improves upon purely greedy constructions.

\subsubsection{KaHyPar}
KaHyPar is a multilevel hypergraph partitioner \cite{schlag2023high}.
It applies hypergraph coarsening,
initial partitioning of the reduced instance,
and uncoarsening with refinement.
It optimises balanced $k$-way cut objectives
using configurable strategies.
KaHyPar represents state-of-the-art multilevel partitioning
and serves as a strong originical benchmark.

\subsubsection{Zoltan PHG}
Zoltan PHG (Parallel HyperGraph) \cite{devine2006parallel} is designed for scalable
parallel hypergraph partitioning.
It employs distributed-memory refinement strategies
and is commonly used in high-performance computing settings.
In our framework, hypergraphs are serialized
and passed to the external solver,
which returns the cut value.
Zoltan provides a production-grade HPC comparison point.

\subsubsection{Evolutionary Algorithm (EA)}
Our evolutionary algorithm is a population-based
hypergraph partitioner.
Each individual encodes a vertex-to-block assignment.
The initial population includes balanced round-robin,
greedy, and random balanced assignments.
Fitness is defined as
\[
f = \text{cut\_edges} + n \cdot \sum_{b=1}^{k} \max\!\left(0,\, |b| - \text{max\_block}\right),
\]
where $|b|$ is the number of vertices in block $b$.
Selection is performed via tournament selection.
Crossover uses uniform partition crossover
followed by repair to restore balance.
Mutation randomly reassigns vertices under capacity constraints.
Elitism preserves the best individuals across generations.
Unlike other methods, the EA can therefore escape local minima,
at the cost of increased runtime. 

\subsection{Circuits}
We evaluate these partitioners across three distinct origins of circuits.

\subsubsection{Real}
Circuits derived from well-known quantum algorithms.
This origin includes circuits from the 
MQT Benchmark suite \cite{quetschlich2023mqt} 
and Quipper algorithm implementations \cite{green2013quipper}.
These circuits exhibit real structured entanglement patterns, 
causal layering, and connectivity 
that reflects realistic implementations.

\subsubsection{Random}
In contrast we consider a suite of random circuits, 
generated using
Qiskit's \cite{abraham2019qiskit} random circuit function generator (which generates uniform gate sampling)
and it's graph-based random circuits generator (which generated connectivity-driven randomness) 
implemented on randomly sampled interaction graphs that vary across runs.

\subsubsection{Generated}

Finally, we evaluate circuits produced by QGen \cite{mao2025q},
a high-level parameterized quantum circuit generator.
QGen programmatically instantiates well-known quantum algorithms
(e.g., QFT, QPE, Grover, QAOA, VQE, VQC)
with algorithm-specific generation parameters beyond qubit count.
These parameters (pre-set in the suite) control oracle structure, entanglement topology,
variational depth, repetition counts, and problem size,
generating scalable generators of structurally consistent circuits.
Unlike random circuits, these workloads preserve algorithmic structure and application-driven connectivity.
However, unlike real circuits used in this study, QGen circuits are not fixed historical implementations extracted from benchmark repositories.
Instead, they are systematically constructed, parameterized generators derived from algorithmic templates.
This distinction is important: real circuits reflect concrete design decisions, compilation artifacts, and implementation idiosyncrasies,
whereas QGen provides controlled structural scaling without inheriting such instance-specific characteristics.
We include QGen circuits to investigate whether structured, parameterized circuit generators can serve as scalable stress-test workloads for partitioning heuristics.
Current benchmark suites typically remain below 200 qubits, limiting the study of heuristic behavior in large distributed regimes.
Generators such as QGen could enable the systematic scaling of structured circuits while preserving their semantic and architectural properties,
thereby offering a controllable bridge between fixed real workloads and structureless random ensembles 
for DQC partitioner training and testing.

\subsection{Experimental Setup and Analysis}

We evaluate all partitioning strategies across a range of
fully connected $k$-QPU network topologies,
with
$k \in \{2, \dots, 10\}$.

A minimum of $5$ qubits per QPU is enforced,
so only configurations satisfying
$n \geq 5k$
are considered.
Circuits exceeding $130$ qubits are excluded
to ensure comparable testing across all circuit origins.
Additionally, circuits containing more than $20{,}000$
multi-qubit gates are skipped to avoid pathological cases.

For each circuit and each valid $k$,
we construct the corresponding hypergraph and
apply every partitioning strategy independently.

We assume a fully connected inter-QPU network.
Thus, any cross-partition hyperedge
can in principle be implemented
via a single non-local gate primitive,
avoiding routing.
This is a standard simplification for near-term DQC settings where direct inter-QPU channels can be established via entanglement swapping.
All presented results are averaged accross all 9 network topologies.

For each $(\text{circuit}, k)$ pair:
\begin{enumerate}
    \item The circuit is parsed into a hypergraph representation.
    \item The partitioner produces a balanced $k$-way assignment.
    \item The number of cut hyperedges is recorded as the cut cost.
\end{enumerate}

Results are aggregated across:
\begin{itemize}
    \item Circuit origin (Real, Random, Generated),
    \item Circuit size (n),
    \item Number of QPUs / partitions (k),
    \item Partitioning strategy.
\end{itemize}

This framework enables the evaluation of scaling behaviour under increasing partition counts,
and aims to isolate the effect of circuit structure on partitioning performance.
Each (strategy, $k$) configuration is evaluated once across 6,508 circuits; 
cross-circuit variability within each ensemble is captured through distributional analysis. 
While several strategies are non-deterministic, outputs were qualitatively consistent across instances. 
Distributional differences between circuit origins are validated using pairwise Mann-Whitney U tests \cite{mann1947test} on normalised cut costs (divided by qubit count),
with effect sizes reported as rank-biserial correlations~$r$.

The central question is whether circuit ensemble choice induces distortion in perceived partitioner performance. We operationalize distortion along four axes:
\begin{itemize}
    \item Cut cost (equivalently, \emph{Communication cost}): the number of cut hyperedges (those whose vertices span more than one partition), each of which requires consuming a non-local EPR pair in the distributed setting,
    \item Scaling behaviour across QPU counts and circuit widths,
    \item Relative ranking of partitioning strategies,
    \item Variability of results across instances.
\end{itemize}

A benchmarking ensemble is considered less distorted
if it preserves the qualitative ordering, scaling trends,
and dispersion characteristics observed on real circuits.

\section{Results}

\begin{figure}
    \centering
    \def\svgwidth{0.5\textwidth}%
    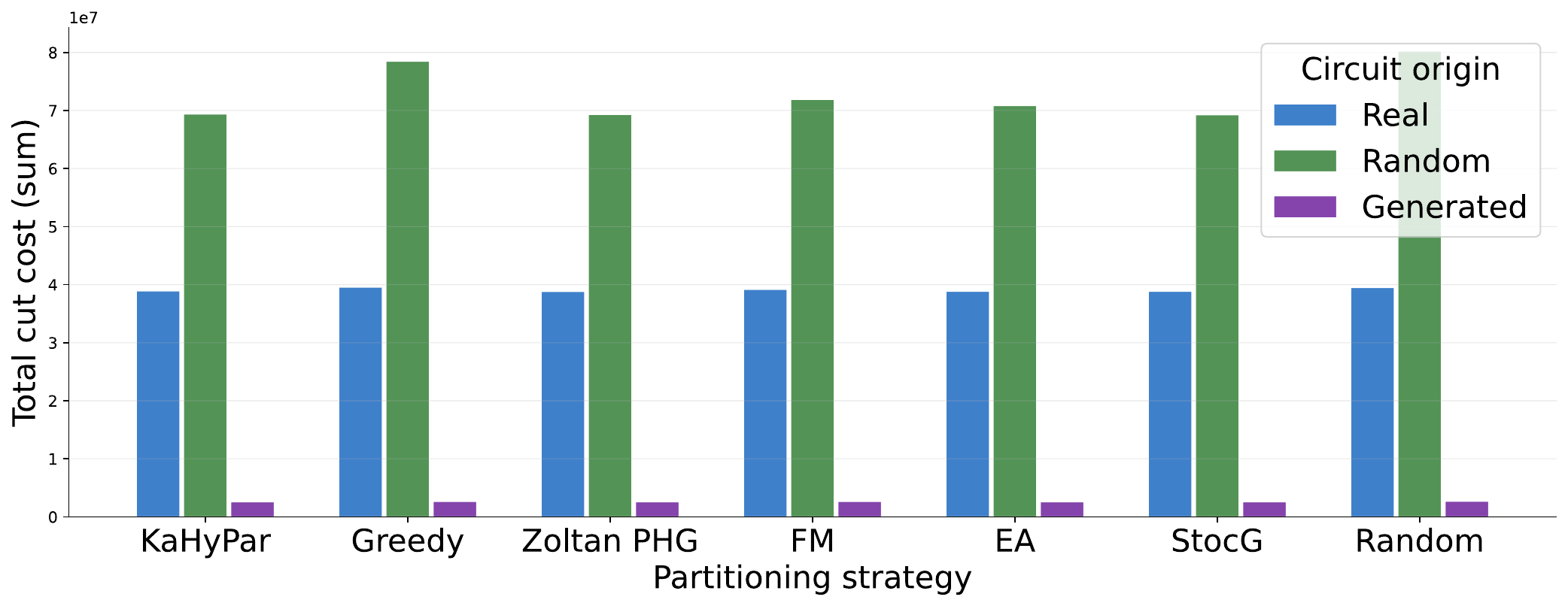
\caption{Total cut cost produced by each partitioning strategy across real, generated, and random circuit origins.
}
\label{fig:variancecost}
\vspace{-0.5cm} \end{figure}

The first clear observation from the results is that the origin of a circuit affects partitioner performance.
In Figure \ref{fig:variancecost}, the total accumulated cut cost across all partitioners and circuit configurations is substantially larger for random circuits than for real circuits, which are in turn larger than generated circuits. 
The separation between these three origins is consistent across strategies.

\begin{figure}
    \centering
    \def\svgwidth{0.5\textwidth}%
    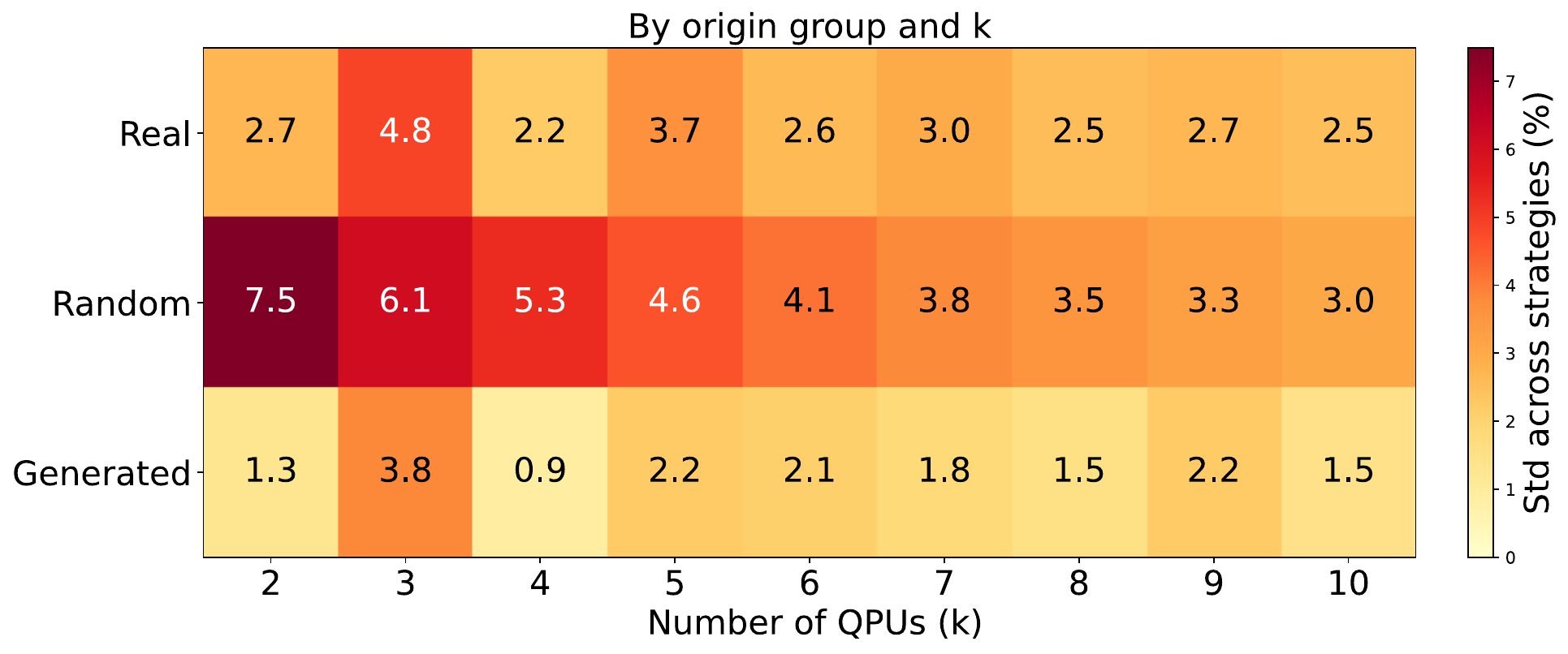
\caption{Heatmap of partitioning performance spread across circuit origins and strategies.
}
\label{fig:heatmat}
\vspace{-0.5cm} \end{figure}

Additionally, Figure \ref{fig:heatmat} shows that the variance across partitioning strategies and QPU sizes is considerably higher for random circuits. 
In contrast, real and generated circuits exhibit smaller variances, 
with tighter clustering of results across configurations.

\begin{figure*}
    \centering
    \def\svgwidth{0.8\textwidth}%
    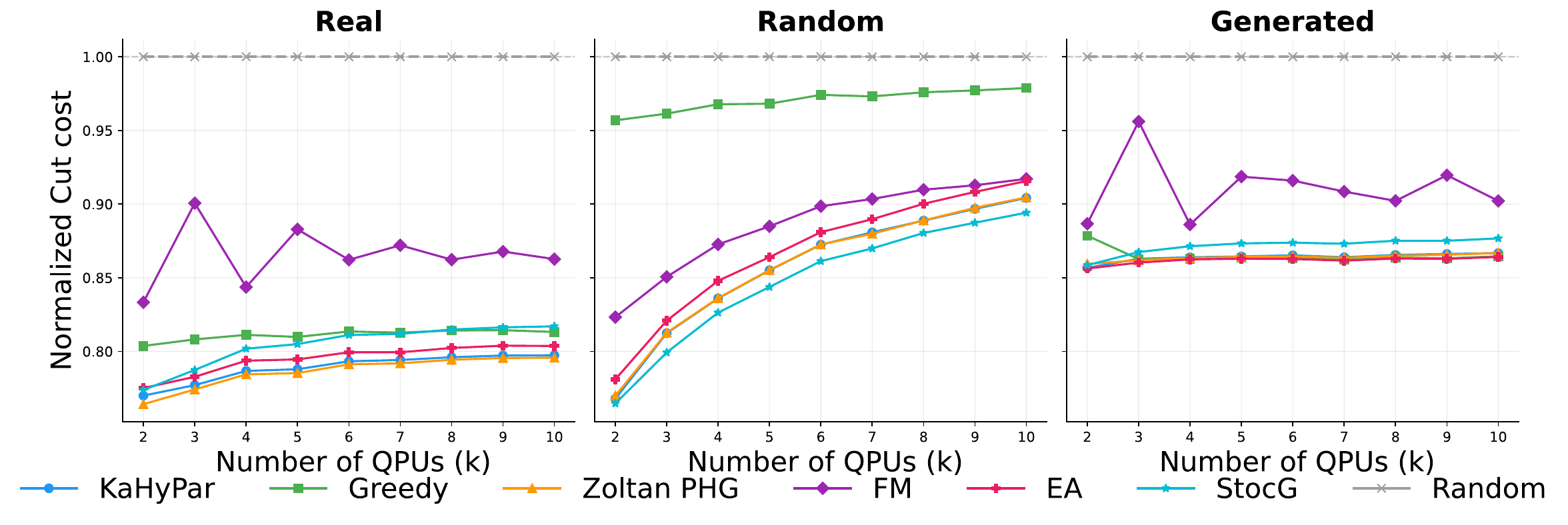
\caption{Normalised cut cost (relative to the random assignment baseline) as a function of the number of QPUs. 
}
\label{fig:byqpus}
\vspace{-0.7 cm}
\end{figure*}

\begin{figure*}
    \centering
    \def\svgwidth{0.8\textwidth}%
    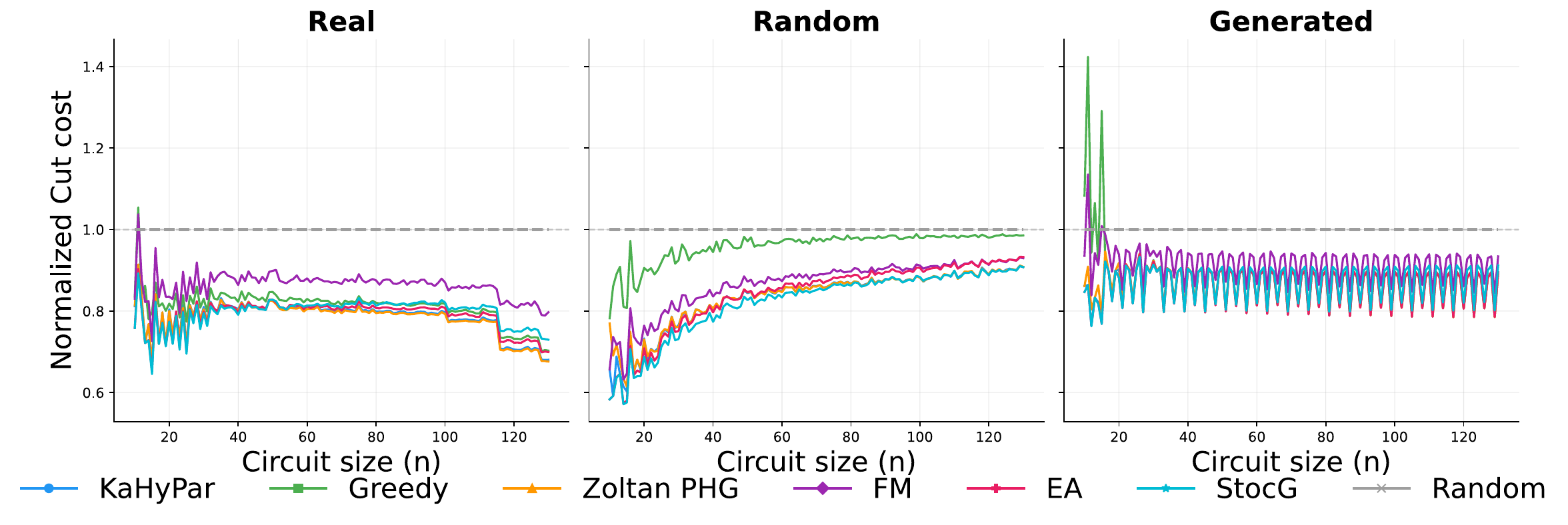
\caption{Normalised cut cost (relative to the random assignment baseline) as a function of circuit size 
(number of qubits). 
}
\label{fig:bysize}
\vspace{-0.1 cm}
\end{figure*}

Beyond this aggregate cost and variance mismatch, we observe distortion in the relative ranking of partitioning strategies across circuit origins. 
Figures \ref{fig:byqpus} and \ref{fig:bysize} show the normalised cut cost relative to 
the random partitioner (baseline), across QPU counts and circuit sizes respectively.

Across QPU counts (Figure \ref{fig:byqpus}), FM consistently incurs higher normalised cost for both real and generated circuits. For random circuits, however, the greedy strategy exhibits the highest relative cost. Other techniques interchange their relative positions depending on circuit origin, but the ordering remains more stable between real and generated circuits than for random ones.

When scaling with circuit size (Figure \ref{fig:bysize}), distinct behaviours emerge across circuit origins.
Real circuits exhibit high cut cost at small sizes, which then stabilises and gradually tapers off beyond approximately 100 qubits across techniques. 
FM again yields the highest cost in this category. 
Generated circuits follow a broadly similar trend, but with significantly higher variance across sizes, particularly below 20 qubits. 
They do not display the same pronounced stabilisation as real circuits at larger scales. 
Random circuits show a steady increase in cost with circuit size, progressively approaching the random baseline. 
This behaviour differs from that observed in real circuits, where cost stabilises rather than grows. 
Notably, random circuits are the only origin for which no partitioning technique performs worse than the random baseline. 
For both real and generated circuits, some techniques underperform the baseline at small sizes.

\begin{figure}
    \centering
    \def\svgwidth{0.5\textwidth}%
    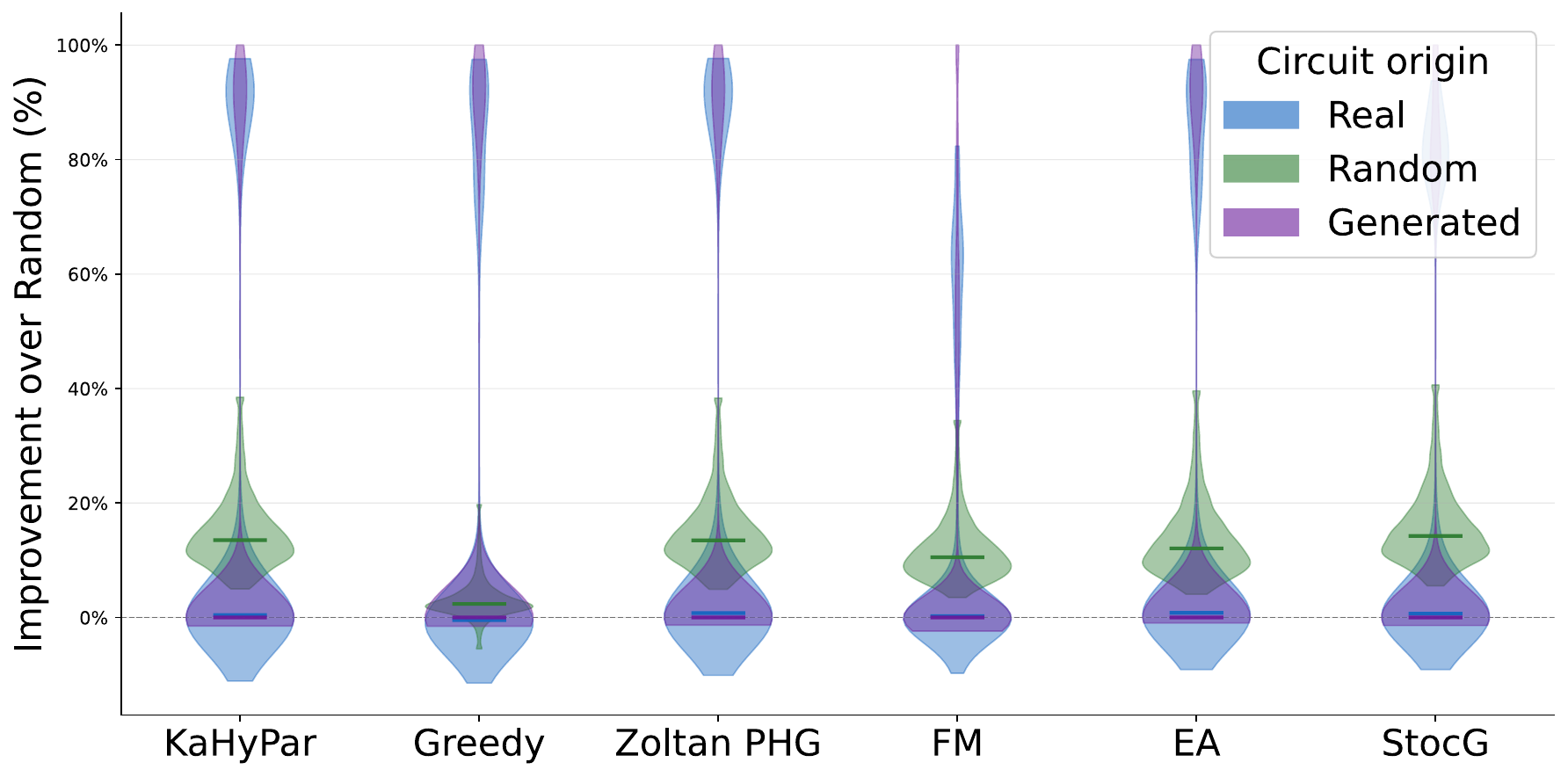
\caption{Violin plots showing the distribution of total cut cost across partitioning strategies for real, generated, and random circuit origins.
The shape and spread of each violin highlight how circuit origin influences performance variability per partitioner.}
\label{fig:violing}
\vspace{-0.5cm} \end{figure}

Figure \ref{fig:violing} shows the aggregated cut cost distributions across all network sizes, partitioners, and circuit sizes. 
Generated circuits closely track the distribution of real circuits, exhibiting substantial overlap in both spread and central tendency. 
Notably, unlike real circuits, 
random circuits display broader dispersion and weaker overlap with the other two origins.

Ranking distortion is quantified via the Spearman rank correlation \cite{spearman1961proof} of strategy rankings relative to real circuits ($\rho = 1$ indicates perfect agreement).
Generated circuits achieve $\rho = 0.771$; random circuits only $\rho = 0.543$, confirming substantially greater ranking distortion for random ensembles.
Mann-Whitney U tests on normalised cut costs confirm that real and random circuit distributions are statistically distinguishable ($p < 0.001$) but with negligible practical effect size ($r = 0.023$): the two ensembles present similar aggregate difficulty levels.
The pronounced ranking divergence ($\rho = 0.543$) therefore reflects a structural reordering of strategies rather than a shift in overall cost magnitude.
Real and generated circuits differ substantially in cut cost ($r = -0.83$, $p < 0.001$), consistent with the markedly lower two-qubit gate density of generated circuits.

The normalised scaling plots (Figures \ref{fig:byqpus} and \ref{fig:bysize}) and the distributional violin plot (Figure \ref{fig:violing}) offer complementary views of the same data.
The scaling plots isolate structural growth and ranking behaviour across QPU counts and circuit widths, abstracting away absolute magnitude.
The violin plot exposes dispersion and overlap across instances, revealing stability differences not visible in averaged trends.
Taken together, they show that random circuits induce distortion across all four identified axes: cut cost, scaling behaviour, variability, and strategy ordering.

Tables~\ref{tab:rank_k} and \ref{tab:rank_size} report explicit ranking positions rather than aggregate statistics.
For real circuits, Zoltan PHG most frequently occupies the top position across QPU counts, with KaHyPar second and FM consistently last.
The ordering shifts substantially across circuit origins: StocG leads for random circuits, where Greedy collapses to last; EA leads for generated circuits, where the remaining strategies form a compressed pack.
Strategies that rank near the bottom for one origin do not remain there across others: FM ranks 5th (not last) for random circuits, and Greedy, last for random, performs mid-table for real and generated workloads.
This instability extends to the top of the ranking, where the leading strategy changes entirely across circuit origins.
Since practitioners face deployment decisions precisely among the top-tier strategies, this is the most consequential form of distortion induced by benchmark choice.

\begin{table*}
\centering
\resizebox{2\columnwidth}{!}{
\begin{filecontents*}{tablerankingsbyk.csv}
origin_group,k,rank_1,rank_2,rank_3,rank_4,rank_5,rank_6
Real,2,\ZoltanPHG (+23.6\%),\KaHyPar (+23.0\%),\StocG (+22.6\%),\EA (+22.5\%),\Greedy (+19.6\%),\FM (+16.7\%)
Real,3,\ZoltanPHG (+22.6\%),\KaHyPar (+22.3\%),\EA (+21.7\%),\StocG (+21.3\%),\Greedy (+19.2\%),\FM (+9.9\%)
Real,4,\ZoltanPHG (+21.6\%),\KaHyPar (+21.3\%),\EA (+20.6\%),\StocG (+19.8\%),\Greedy (+18.9\%),\FM (+15.6\%)
Real,5,\ZoltanPHG (+21.5\%),\KaHyPar (+21.2\%),\EA (+20.5\%),\StocG (+19.5\%),\Greedy (+19.0\%),\FM (+11.7\%)
Real,6,\ZoltanPHG (+20.9\%),\KaHyPar (+20.7\%),\EA (+20.1\%),\StocG (+18.9\%),\Greedy (+18.6\%),\FM (+13.8\%)
Real,7,\ZoltanPHG (+20.8\%),\KaHyPar (+20.6\%),\EA (+20.1\%),\StocG (+18.8\%),\Greedy (+18.7\%),\FM (+12.8\%)
Real,8,\ZoltanPHG (+20.6\%),\KaHyPar (+20.4\%),\EA (+19.8\%),\Greedy (+18.6\%),\StocG (+18.5\%),\FM (+13.8\%)
Real,9,\ZoltanPHG (+20.5\%),\KaHyPar (+20.3\%),\EA (+19.6\%),\Greedy (+18.6\%),\StocG (+18.4\%),\FM (+13.2\%)
Real,10,\ZoltanPHG (+20.4\%),\KaHyPar (+20.3\%),\EA (+19.6\%),\Greedy (+18.7\%),\StocG (+18.3\%),\FM (+13.7\%)
Random,2,\StocG (+23.5\%),\KaHyPar (+23.2\%),\ZoltanPHG (+23.0\%),\EA (+21.9\%),\FM (+17.7\%),\Greedy (+4.3\%)
Random,3,\StocG (+20.1\%),\KaHyPar (+18.8\%),\ZoltanPHG (+18.7\%),\EA (+17.9\%),\FM (+14.9\%),\Greedy (+3.9\%)
Random,4,\StocG (+17.4\%),\KaHyPar (+16.4\%),\ZoltanPHG (+16.4\%),\EA (+15.2\%),\FM (+12.7\%),\Greedy (+3.2\%)
Random,5,\StocG (+15.6\%),\KaHyPar (+14.5\%),\ZoltanPHG (+14.5\%),\EA (+13.6\%),\FM (+11.5\%),\Greedy (+3.2\%)
Random,6,\StocG (+13.9\%),\KaHyPar (+12.8\%),\ZoltanPHG (+12.7\%),\EA (+11.9\%),\FM (+10.1\%),\Greedy (+2.6\%)
Random,7,\StocG (+13.0\%),\ZoltanPHG (+12.0\%),\KaHyPar (+11.9\%),\EA (+11.0\%),\FM (+9.7\%),\Greedy (+2.7\%)
Random,8,\StocG (+12.0\%),\KaHyPar (+11.1\%),\ZoltanPHG (+11.1\%),\EA (+10.0\%),\FM (+9.0\%),\Greedy (+2.4\%)
Random,9,\StocG (+11.3\%),\KaHyPar (+10.3\%),\ZoltanPHG (+10.3\%),\EA (+9.2\%),\FM (+8.7\%),\Greedy (+2.3\%)
Random,10,\StocG (+10.6\%),\KaHyPar (+9.6\%),\ZoltanPHG (+9.5\%),\EA (+8.4\%),\FM (+8.3\%),\Greedy (+2.1\%)
Generated,2,\EA (+14.4\%),\KaHyPar (+14.3\%),\StocG (+14.1\%),\ZoltanPHG (+14.1\%),\Greedy (+12.2\%),\FM (+11.3\%)
Generated,3,\EA (+14.0\%),\ZoltanPHG (+13.8\%),\Greedy (+13.7\%),\KaHyPar (+13.7\%),\StocG (+13.3\%),\FM (+4.4\%)
Generated,4,\EA (+13.8\%),\Greedy (+13.7\%),\ZoltanPHG (+13.7\%),\KaHyPar (+13.6\%),\StocG (+12.9\%),\FM (+11.4\%)
Generated,5,\EA (+13.7\%),\Greedy (+13.7\%),\ZoltanPHG (+13.6\%),\KaHyPar (+13.6\%),\StocG (+12.7\%),\FM (+8.1\%)
Generated,6,\EA (+13.7\%),\Greedy (+13.7\%),\ZoltanPHG (+13.6\%),\KaHyPar (+13.5\%),\StocG (+12.6\%),\FM (+8.4\%)
Generated,7,\EA (+13.8\%),\Greedy (+13.8\%),\ZoltanPHG (+13.6\%),\KaHyPar (+13.6\%),\StocG (+12.7\%),\FM (+9.2\%)
Generated,8,\EA (+13.7\%),\Greedy (+13.6\%),\ZoltanPHG (+13.5\%),\KaHyPar (+13.5\%),\StocG (+12.5\%),\FM (+9.8\%)
Generated,9,\EA (+13.7\%),\Greedy (+13.7\%),\ZoltanPHG (+13.4\%),\KaHyPar (+13.4\%),\StocG (+12.5\%),\FM (+8.0\%)
Generated,10,\EA (+13.6\%),\Greedy (+13.6\%),\KaHyPar (+13.3\%),\ZoltanPHG (+13.3\%),\StocG (+12.3\%),\FM (+9.8\%)
\end{filecontents*}
\pgfplotstabletypeset[
    col sep=comma,
    string type,
    trim cells=true,
    columns={origin_group,k,rank_1,rank_2,rank_3,rank_4,rank_5,rank_6},
    columns/origin_group/.style={column name=Circuit Class},
    columns/k/.style={column name=$k$},
    columns/rank_1/.style={column name=1st},
    columns/rank_2/.style={column name=2nd},
    columns/rank_3/.style={column name=3rd},
    columns/rank_4/.style={column name=4th},
    columns/rank_5/.style={column name=5th},
    columns/rank_6/.style={column name=6th},
    every head row/.style={
        before row=\toprule,
        after row=\midrule,
        font=\bfseries
    },
    every last row/.style={
        after row=\bottomrule
    }
]{tablerankingsbyk.csv}
}
\caption{Partitioner ranking by circuit origin and $k$.}
\label{tab:rank_k}
\vspace{-0.6 cm}
\end{table*}

\begin{table*}
\centering
\resizebox{2\columnwidth}{!}{
\begin{filecontents*}{tablerankingsbyqubitbin.csv}
origin_group,qubit_bin,rank_1,rank_2,rank_3,rank_4,rank_5,rank_6
Real,0–14,\StocG (+21.7\%),\EA (+21.4\%),\KaHyPar (+19.4\%),\ZoltanPHG (+18.9\%),\Greedy (+11.9\%),\FM (+11.8\%)
Real,15–29,\StocG (+24.8\%),\EA (+24.3\%),\ZoltanPHG (+23.1\%),\KaHyPar (+23.0\%),\Greedy (+18.3\%),\FM (+15.1\%)
Real,30–44,\StocG (+19.7\%),\EA (+19.3\%),\ZoltanPHG (+19.1\%),\KaHyPar (+18.9\%),\Greedy (+16.5\%),\FM (+11.9\%)
Real,45–59,\ZoltanPHG (+19.1\%),\KaHyPar (+18.9\%),\StocG (+18.7\%),\EA (+18.7\%),\Greedy (+16.9\%),\FM (+12.0\%)
Real,60–74,\ZoltanPHG (+19.9\%),\KaHyPar (+19.6\%),\EA (+19.1\%),\StocG (+18.5\%),\Greedy (+17.7\%),\FM (+12.6\%)
Real,75–89,\ZoltanPHG (+20.3\%),\KaHyPar (+20.0\%),\EA (+19.1\%),\StocG (+18.1\%),\Greedy (+18.0\%),\FM (+12.6\%)
Real,90–104,\ZoltanPHG (+21.2\%),\KaHyPar (+20.8\%),\EA (+19.7\%),\Greedy (+18.8\%),\StocG (+18.3\%),\FM (+13.3\%)
Real,105–119,\ZoltanPHG (+24.0\%),\KaHyPar (+23.7\%),\EA (+22.4\%),\Greedy (+21.5\%),\StocG (+20.3\%),\FM (+14.9\%)
Real,120–134,\ZoltanPHG (+30.3\%),\KaHyPar (+30.0\%),\EA (+28.1\%),\Greedy (+27.4\%),\StocG (+25.2\%),\FM (+19.1\%)
Random,0–14,\StocG (+39.4\%),\EA (+39.4\%),\KaHyPar (+36.1\%),\FM (+30.7\%),\ZoltanPHG (+30.2\%),\Greedy (+14.9\%)
Random,15–29,\StocG (+31.0\%),\EA (+29.4\%),\KaHyPar (+28.0\%),\ZoltanPHG (+27.8\%),\FM (+22.9\%),\Greedy (+9.2\%)
Random,30–44,\StocG (+22.2\%),\EA (+20.3\%),\KaHyPar (+19.7\%),\ZoltanPHG (+19.7\%),\FM (+16.1\%),\Greedy (+4.9\%)
Random,45–59,\StocG (+17.5\%),\KaHyPar (+15.9\%),\ZoltanPHG (+15.8\%),\EA (+15.6\%),\FM (+12.9\%),\Greedy (+3.5\%)
Random,60–74,\StocG (+15.2\%),\ZoltanPHG (+14.3\%),\KaHyPar (+14.3\%),\EA (+13.2\%),\FM (+11.4\%),\Greedy (+2.6\%)
Random,75–89,\StocG (+13.4\%),\ZoltanPHG (+13.0\%),\KaHyPar (+13.0\%),\EA (+11.4\%),\FM (+10.0\%),\Greedy (+2.1\%)
Random,90–104,\ZoltanPHG (+12.2\%),\StocG (+12.2\%),\KaHyPar (+12.1\%),\EA (+10.1\%),\FM (+9.3\%),\Greedy (+1.8\%)
Random,105–119,\StocG (+11.1\%),\ZoltanPHG (+11.1\%),\KaHyPar (+11.1\%),\EA (+8.8\%),\FM (+8.6\%),\Greedy (+1.7\%)
Random,120–134,\StocG (+10.0\%),\KaHyPar (+9.9\%),\ZoltanPHG (+9.9\%),\FM (+7.6\%),\EA (+7.5\%),\Greedy (+1.5\%)
Generated,0–14,\StocG (+17.7\%),\EA (+17.7\%),\KaHyPar (+17.3\%),\ZoltanPHG (+14.7\%),\FM (+5.2\%),\Greedy (-7.9\%)
Generated,15–29,\StocG (+12.6\%),\EA (+12.5\%),\KaHyPar (+12.0\%),\ZoltanPHG (+11.5\%),\Greedy (+9.7\%),\FM (+6.6\%)
Generated,30–44,\StocG (+12.2\%),\EA (+12.2\%),\Greedy (+12.1\%),\ZoltanPHG (+11.8\%),\KaHyPar (+11.8\%),\FM (+7.7\%)
Generated,45–59,\EA (+13.6\%),\Greedy (+13.5\%),\ZoltanPHG (+13.3\%),\KaHyPar (+13.3\%),\StocG (+13.1\%),\FM (+8.7\%)
Generated,60–74,\EA (+13.8\%),\Greedy (+13.7\%),\ZoltanPHG (+13.7\%),\KaHyPar (+13.6\%),\StocG (+13.0\%),\FM (+9.1\%)
Generated,75–89,\EA (+14.2\%),\Greedy (+14.1\%),\ZoltanPHG (+14.0\%),\KaHyPar (+14.0\%),\StocG (+13.0\%),\FM (+9.3\%)
Generated,90–104,\EA (+14.2\%),\ZoltanPHG (+14.2\%),\Greedy (+14.2\%),\KaHyPar (+14.1\%),\StocG (+12.8\%),\FM (+9.3\%)
Generated,105–119,\EA (+14.4\%),\Greedy (+14.4\%),\ZoltanPHG (+14.4\%),\KaHyPar (+14.3\%),\StocG (+12.8\%),\FM (+9.4\%)
Generated,120–134,\EA (+14.7\%),\ZoltanPHG (+14.7\%),\Greedy (+14.6\%),\KaHyPar (+14.6\%),\StocG (+13.0\%),\FM (+9.8\%)
\end{filecontents*}
\pgfplotstabletypeset[
    col sep=comma,
    string type,
    trim cells=true,
    columns={origin_group,qubit_bin,rank_1,rank_2,rank_3,rank_4,rank_5,rank_6},
    columns/origin_group/.style={column name=Circuit Class},
    columns/qubit_bin/.style={column name=n qubits},
    columns/rank_1/.style={column name=1st},
    columns/rank_2/.style={column name=2nd},
    columns/rank_3/.style={column name=3rd},
    columns/rank_4/.style={column name=4th},
    columns/rank_5/.style={column name=5th},
    columns/rank_6/.style={column name=6th},
    every head row/.style={
        before row=\toprule,
        after row=\midrule,
        font=\bfseries
    },
    every last row/.style={
        after row=\bottomrule
    }
]{tablerankingsbyqubitbin.csv}
}
\caption{Partitioner ranking by circuit origin and size.
Results are reported in 15-qubit increments.}
\label{tab:rank_size}
\vspace{-0.8 cm}
\end{table*}

\section{Discussion and Future Work}
Our results show that circuit origin alters partitioning behaviour.
Random circuits not only produce higher absolute cut costs, but also alter the relative performance ordering of partitioning strategies. 
Methods that appear strong under random workloads do not retain this position for real circuits, and vice versa.
This ranking instability (reflected in the large variance across methods and QPU counts) indicates that random circuits 
are not only a harder more compute consuming benchmark, but they induce a different evaluation regime, 
leading to incorrect assumptions about the relative merit of partitioning strategies. 
Evaluating DQC partitioners on random circuits therefore risks selecting strategies that are suboptimal for realistic workloads.

One might attribute the higher absolute cut costs of random circuits to higher two-qubit gate density.
The data does not support this.
Through our tests the two random generators produced markedly different gate fractions (23\% and 66\%) but both yield elevated cut costs relative to real circuits.
Real MQT and Quipper circuits, with the highest density of all ensembles (76\% and 66.2\%), consistently incured lower costs than either random generator.
Generated (QGen) circuits present very low two-qubit gate density (5\%), explaining their lower absolute cut costs. 
Despite this their ranking agreement with real workloads remained substantially higher ($\rho = 0.771$ vs $0.543$), 
implying that ranking fidelity reflects more than raw gate count.
We therefore conjecture that the distinguishing structural property is qubit \emph{interaction topology} (the pattern of which qubits interact and how, encoded in the hyperedge structure) rather than multi-qubit gate count.

Overall, generated circuits produce substantially lower cut costs than random circuits and exhibit higher rank agreement with real workloads. 
However, in several configurations, their top-ranked strategy differs from that of real circuits.

Despite this, overall, generated circuits do provide a closer approximation to real workloads than random ensembles at the level of aggregate partitioning behaviour, 
while consistently inducing lower cut costs. 
This places them in a structurally informed but comparatively less demanding regime, 
making them computationally cheaper to evaluate and therefore well suited for rapid prototyping and early-stage testing of partitioning strategies. 
Given the limited size and diversity of publicly available quantum benchmarks, 
generated circuits offer a scalable intermediate setting between dense random stress tests and scarce large-scale real applications, 
preserving several partitioning-relevant characteristics.

It is important to note that the QGen circuits used in this study are not AI-generated in the sense of learned generative models. 
They are structured, programmatically constructed implementations designed to reflect application-informed characteristics. 
Our results therefore do not demonstrate that machine learning–based generators can reproduce realistic workloads.

However, the partial structural agreement observed between generated and real circuits indicates that principled workload-aware construction can approximate certain partitioning-relevant properties of real applications. 
This motivates future investigation into whether learning–based generative models could be explicitly trained to preserve interaction topologies relevant for distributed quantum computing. 
If successful, such approaches could provide a scalable and structurally grounded alternative to random ensembles for evaluating DQC scalability.

Finally, within real circuits, strategy rankings shift with QPU count and circuit size, indicating that aggregate rankings alone are insufficient to characterise partitioner performance comprehensively.
Future evaluation frameworks should account for circuit-specific structural properties and extend beyond the fully-connected network model to incorporate routing costs, QPUs with varying qubit counts, and other constraints of realistic quantum networks.

Code and data are available at \href{https://github.com/grageragarces/Random-circuits-for-DQC}{github.com/grageragarces/Random-circuits-for-DQC}.
The circuit benchmarking suite is available upon request.

\vspace{-0.2 cm}

\bibliographystyle{IEEEtran}
\bibliography{references}

\end{document}

%% file: svg-inkscape/fig7_total_cut_by_strategy_svg-tex.pdf_tex
\begingroup%
  \makeatletter%
  \providecommand\color[2][]{%
    \errmessage{(Inkscape) Color is used for the text in Inkscape, but the package 'color.sty' is not loaded}%
    \renewcommand\color[2][]{}%
  }%
  \providecommand\transparent[1]{%
    \errmessage{(Inkscape) Transparency is used (non-zero) for the text in Inkscape, but the package 'transparent.sty' is not loaded}%
    \renewcommand\transparent[1]{}%
  }%
  \providecommand\rotatebox[2]{#2}%
  \newcommand*\fsize{\dimexpr\f@size pt\relax}%
  \newcommand*\lineheight[1]{\fontsize{\fsize}{#1\fsize}\selectfont}%
  \ifx\svgwidth\undefined%
    \setlength{\unitlength}{999.96875bp}%
    \ifx\svgscale\undefined%
      \relax%
    \else%
      \setlength{\unitlength}{\unitlength * \real{\svgscale}}%
    \fi%
  \else%
    \setlength{\unitlength}{\svgwidth}%
  \fi%
  \global\let\svgwidth\undefined%
  \global\let\svgscale\undefined%
  \makeatother%
  \begin{picture}(1,0.38688303)%
    \lineheight{1}%
    \setlength\tabcolsep{0pt}%
    \put(0,0){\includegraphics[width=\unitlength,page=1]{fig7_total_cut_by_strategy_svg-tex.pdf}}%
  \end{picture}%
\endgroup%

%% file: svg-inkscape/fig4_technique_spread_svg-tex.pdf_tex
\begingroup%
  \makeatletter%
  \providecommand\color[2][]{%
    \errmessage{(Inkscape) Color is used for the text in Inkscape, but the package 'color.sty' is not loaded}%
    \renewcommand\color[2][]{}%
  }%
  \providecommand\transparent[1]{%
    \errmessage{(Inkscape) Transparency is used (non-zero) for the text in Inkscape, but the package 'transparent.sty' is not loaded}%
    \renewcommand\transparent[1]{}%
  }%
  \providecommand\rotatebox[2]{#2}%
  \newcommand*\fsize{\dimexpr\f@size pt\relax}%
  \newcommand*\lineheight[1]{\fontsize{\fsize}{#1\fsize}\selectfont}%
  \ifx\svgwidth\undefined%
    \setlength{\unitlength}{926.69720459bp}%
    \ifx\svgscale\undefined%
      \relax%
    \else%
      \setlength{\unitlength}{\unitlength * \real{\svgscale}}%
    \fi%
  \else%
    \setlength{\unitlength}{\svgwidth}%
  \fi%
  \global\let\svgwidth\undefined%
  \global\let\svgscale\undefined%
  \makeatother%
  \begin{picture}(1,0.41712587)%
    \lineheight{1}%
    \setlength\tabcolsep{0pt}%
    \put(0,0){\includegraphics[width=\unitlength,page=1]{fig4_technique_spread_svg-tex.pdf}}%
  \end{picture}%
\endgroup%

%% file: svg-inkscape/fig6_normalised_cut_vs_k_svg-tex.pdf_tex
\begingroup%
  \makeatletter%
  \providecommand\color[2][]{%
    \errmessage{(Inkscape) Color is used for the text in Inkscape, but the package 'color.sty' is not loaded}%
    \renewcommand\color[2][]{}%
  }%
  \providecommand\transparent[1]{%
    \errmessage{(Inkscape) Transparency is used (non-zero) for the text in Inkscape, but the package 'transparent.sty' is not loaded}%
    \renewcommand\transparent[1]{}%
  }%
  \providecommand\rotatebox[2]{#2}%
  \newcommand*\fsize{\dimexpr\f@size pt\relax}%
  \newcommand*\lineheight[1]{\fontsize{\fsize}{#1\fsize}\selectfont}%
  \ifx\svgwidth\undefined%
    \setlength{\unitlength}{1128.71252441bp}%
    \ifx\svgscale\undefined%
      \relax%
    \else%
      \setlength{\unitlength}{\unitlength * \real{\svgscale}}%
    \fi%
  \else%
    \setlength{\unitlength}{\svgwidth}%
  \fi%
  \global\let\svgwidth\undefined%
  \global\let\svgscale\undefined%
  \makeatother%
  \begin{picture}(1,0.33503381)%
    \lineheight{1}%
    \setlength\tabcolsep{0pt}%
    \put(0,0){\includegraphics[width=\unitlength,page=1]{fig6_normalised_cut_vs_k_svg-tex.pdf}}%
  \end{picture}%
\endgroup%

%% file: svg-inkscape/fig6_normalised_cut_vs_k_v2_svg-tex.pdf_tex
\begingroup%
  \makeatletter%
  \providecommand\color[2][]{%
    \errmessage{(Inkscape) Color is used for the text in Inkscape, but the package 'color.sty' is not loaded}%
    \renewcommand\color[2][]{}%
  }%
  \providecommand\transparent[1]{%
    \errmessage{(Inkscape) Transparency is used (non-zero) for the text in Inkscape, but the package 'transparent.sty' is not loaded}%
    \renewcommand\transparent[1]{}%
  }%
  \providecommand\rotatebox[2]{#2}%
  \newcommand*\fsize{\dimexpr\f@size pt\relax}%
  \newcommand*\lineheight[1]{\fontsize{\fsize}{#1\fsize}\selectfont}%
  \ifx\svgwidth\undefined%
    \setlength{\unitlength}{1128.71252441bp}%
    \ifx\svgscale\undefined%
      \relax%
    \else%
      \setlength{\unitlength}{\unitlength * \real{\svgscale}}%
    \fi%
  \else%
    \setlength{\unitlength}{\svgwidth}%
  \fi%
  \global\let\svgwidth\undefined%
  \global\let\svgscale\undefined%
  \makeatother%
  \begin{picture}(1,0.33503381)%
    \lineheight{1}%
    \setlength\tabcolsep{0pt}%
    \put(0,0){\includegraphics[width=\unitlength,page=1]{fig6_normalised_cut_vs_k_v2_svg-tex.pdf}}%
  \end{picture}%
\endgroup%

%% file: svg-inkscape/fig3_distributions_svg-tex.pdf_tex
\begingroup%
  \makeatletter%
  \providecommand\color[2][]{%
    \errmessage{(Inkscape) Color is used for the text in Inkscape, but the package 'color.sty' is not loaded}%
    \renewcommand\color[2][]{}%
  }%
  \providecommand\transparent[1]{%
    \errmessage{(Inkscape) Transparency is used (non-zero) for the text in Inkscape, but the package 'transparent.sty' is not loaded}%
    \renewcommand\transparent[1]{}%
  }%
  \providecommand\rotatebox[2]{#2}%
  \newcommand*\fsize{\dimexpr\f@size pt\relax}%
  \newcommand*\lineheight[1]{\fontsize{\fsize}{#1\fsize}\selectfont}%
  \ifx\svgwidth\undefined%
    \setlength{\unitlength}{856.14532471bp}%
    \ifx\svgscale\undefined%
      \relax%
    \else%
      \setlength{\unitlength}{\unitlength * \real{\svgscale}}%
    \fi%
  \else%
    \setlength{\unitlength}{\svgwidth}%
  \fi%
  \global\let\svgwidth\undefined%
  \global\let\svgscale\undefined%
  \makeatother%
  \begin{picture}(1,0.49523864)%
    \lineheight{1}%
    \setlength\tabcolsep{0pt}%
    \put(0,0){\includegraphics[width=\unitlength,page=1]{fig3_distributions_svg-tex.pdf}}%
  \end{picture}%
\endgroup%

%% file: main.bbl
\begin{thebibliography}{10}
\providecommand{\url}[1]{#1}
\csname url@samestyle\endcsname
\providecommand{\newblock}{\relax}
\providecommand{\bibinfo}[2]{#2}
\providecommand{\BIBentrySTDinterwordspacing}{\spaceskip=0pt\relax}
\providecommand{\BIBentryALTinterwordstretchfactor}{4}
\providecommand{\BIBentryALTinterwordspacing}{\spaceskip=\fontdimen2\font plus
\BIBentryALTinterwordstretchfactor\fontdimen3\font minus
  \fontdimen4\font\relax}
\providecommand{\BIBforeignlanguage}[2]{{%
\expandafter\ifx\csname l@#1\endcsname\relax
\typeout{** WARNING: IEEEtran.bst: No hyphenation pattern has been}%
\typeout{** loaded for the language `#1'. Using the pattern for}%
\typeout{** the default language instead.}%
\else
\language=\csname l@#1\endcsname
\fi
#2}}
\providecommand{\BIBdecl}{\relax}
\BIBdecl

\bibitem{cuomo2020towards}
D.~Cuomo, M.~Caleffi, and A.~S. Cacciapuoti, ``Towards a distributed quantum
  computing ecosystem,'' \emph{IET Quantum Communication}, vol.~1, no.~1, pp.
  3--8, 2020.

\bibitem{barral2025review}
D.~Barral \emph{et~al.}, ``Review of distributed quantum computing: From single
  qpu to high performance quantum computing,'' \emph{Computer Science Review},
  2025.

\bibitem{andres2019automated}
P.~Andres-Martinez and C.~Heunen, ``Automated distribution of quantum circuits
  via hypergraph partitioning,'' \emph{Physical Review A}.

\bibitem{schlag2023high}
S.~Schlag \emph{et~al.}, ``High-quality hypergraph partitioning,'' \emph{ACM
  Journal of Experimental Algorithmics}, 2023.

\bibitem{sundaram2023distributing}
R.~G. Sundaram and H.~Gupta, ``Distributing quantum circuits using
  teleportations,'' in \emph{2023 IEEE International Conference on Quantum
  Software (QSW)}.\hskip 1em plus 0.5em minus 0.4em\relax IEEE, 2023, pp.
  186--192.

\bibitem{sundaram2021efficient}
R.~Sundaram, ``Efficient distribution of quantum circuits,'' 2021.

\bibitem{burt2026multilevel}
F.~Burt, K.-C. Chen, and K.~K. Leung, ``A multilevel framework for partitioning
  quantum circuits,'' \emph{Quantum}, vol.~10, p. 1984, 2026.

\bibitem{escofet2023hungarian}
P.~Escofet, A.~Ovide, C.~G. Almudever, E.~Alarc{\'o}n, and S.~Abadal,
  ``Hungarian qubit assignment for optimized mapping of quantum circuits on
  multi-core architectures,'' \emph{IEEE Computer Architecture Letters},
  vol.~22, no.~2, pp. 161--164, 2023.

\bibitem{bandic2023mapping}
M.~Bandic, L.~Prielinger, J.~N{\"u}{\ss}lein, A.~Ovide, S.~Rodrigo, S.~Abadal,
  H.~Van~Someren, G.~Vardoyan, E.~Alarcon, C.~G. Almudever \emph{et~al.},
  ``Mapping quantum circuits to modular architectures with qubo,'' in
  \emph{2023 IEEE international conference on quantum computing and engineering
  (QCE)}, vol.~1.\hskip 1em plus 0.5em minus 0.4em\relax IEEE, 2023, pp.
  790--801.

\bibitem{arute2019quantum}
F.~Arute, K.~Arya, R.~Babbush, D.~Bacon, J.~C. Bardin, R.~Barends, R.~Biswas,
  S.~Boixo, F.~G. Brandao, D.~A. Buell \emph{et~al.}, ``Quantum supremacy using
  a programmable superconducting processor,'' \emph{nature}, vol. 574, no.
  7779, pp. 505--510, 2019.

\bibitem{fisher2023random}
M.~P. Fisher, V.~Khemani, A.~Nahum, and S.~Vijay, ``Random quantum circuits,''
  \emph{Annual Review of Condensed Matter Physics}, vol.~14, no.~1, pp.
  335--379, 2023.

\bibitem{ccatalyurek2023more}
{\"U}.~{\c{C}}ataly{\"u}rek \emph{et~al.}, ``More recent advances in (hyper)
  graph partitioning,'' \emph{ACM Computing Surveys}, vol.~55, no.~12, pp.
  1--38, 2023.

\bibitem{fiduccia1988linear}
C.~M. Fiduccia and R.~M. Mattheyses, ``A linear-time heuristic for improving
  network partitions,'' in \emph{Papers on Twenty-five years of electronic
  design automation}, 1988, pp. 241--247.

\bibitem{devine2006parallel}
K.~D. Devine, E.~G. Boman, R.~T. Heaphy, R.~H. Bisseling, and U.~V. Catalyurek,
  ``Parallel hypergraph partitioning for scientific computing,'' in
  \emph{Proceedings 20th IEEE International Parallel \& Distributed Processing
  Symposium}.\hskip 1em plus 0.5em minus 0.4em\relax IEEE, 2006, pp. 10--pp.

\bibitem{quetschlich2023mqt}
N.~Quetschlich, L.~Burgholzer, and R.~Wille, ``Mqt bench: Benchmarking software
  and design automation tools for quantum computing,'' \emph{Quantum}, 2023.

\bibitem{green2013quipper}
A.~S. Green, P.~L. Lumsdaine, N.~J. Ross, P.~Selinger, and B.~Valiron,
  ``Quipper: a scalable quantum programming language,'' in \emph{Proceedings of
  the 34th ACM SIGPLAN conference on Programming language design and
  implementation}, 2013, pp. 333--342.

\bibitem{abraham2019qiskit}
G.~Aleksandrowicz \emph{et~al.}, ``Qiskit: An open-source quantum computing
  framework,'' \emph{URL https://doi. org/10.5281/zenodo}, vol. 2562110, 2019.

\bibitem{mao2025q}
Y.~Mao, S.~Shresthamali, and M.~Kondo, ``Q-gen: A parameterized quantum circuit
  generator,'' \emph{TQE}, 2025.

\bibitem{mann1947test}
H.~B. Mann and D.~R. Whitney, ``On a test of whether one of two random
  variables is stochastically larger than the other,'' \emph{The Annals of
  Mathematical Statistics}, vol.~18, no.~1, pp. 50--60, 1947.

\bibitem{spearman1961proof}
C.~Spearman, ``The proof and measurement of association between two things.''
  1961.

\end{thebibliography}
